\RequirePackage[2020-02-02]{latexrelease}
\documentclass[aps,pre,floatfix,twocolumn,nofootinbib]{revtex4-2}
\usepackage{graphicx}
\usepackage{latexsym}
\usepackage{amsmath}
\usepackage{amsfonts}
\usepackage{amssymb}
\usepackage{color}
\usepackage{ulem}
\usepackage[english]{babel}

\parskip=\medskipamount

\newcommand{\be}{\begin{equation}}
\newcommand{\ee}{\end{equation}}
\newcommand{\bea}{\begin{eqnarray}}
\newcommand{\eea}{\end{eqnarray}}

\newcommand{\eq}[1]{(\ref{#1})}
\newcommand{\fig}[1]{Fig.~\ref{#1}}

\newcommand{\eps}{\varepsilon}

\begin{document}

\title{Stretching of a fractal polymer around a disc reveals KPZ scaling}

\author{Kirill E. Polovnikov$^1$, Sergei K. Nechaev$^{2}$, Alexander Y. Grosberg$^3$}

\affiliation{$^1$Skolkovo Institute of Science and Technology, 121205 Moscow, Russia \\ $^2$ LPTMS, Universit\'{e} Paris Saclay,
91405 Orsay Cedex, France 
\\ $^3$Department of Physics and Center for Soft Matter Research, New York University, 726 Broadway, NY 10003, New York, USA}

\begin{abstract}
While stretching of a polymer along a flat surface is hardly different from the classical Pincus problem of pulling chain ends in free space, the role of curved geometry in conformational statistics of the stretched chain is an exciting open question. We use scaling analysis and computer simulations to examine stretching of a fractal polymer chain around a disc in 2D (or a cylinder in 3D) of radius $R$.  We reveal that the typical excursions of the polymer away from the surface and curvature-induced correlation length scale as $\Delta \sim R^{\beta}$ and $S^{*} \sim R^{1/z}$, respectively, with the Kardar-Parisi-Zhang (KPZ) growth $\beta=1/3$ and dynamic exponents $z=3/2$.  Although probability distribution of excursions doesn't belong to KPZ universality class, the KPZ scaling is independent of the fractal dimension of the polymer and, thus, is universal across classical polymer models, e.g. SAW, randomly-branching polymers, crumpled unknotted rings. Additionally, our work establishes a mapping between stretched polymers in curved geometry and Balagurov-Vaks model of random walks among traps.

\end{abstract}

\maketitle


\textit{Introduction.} The phenomenon of stretching a polymer fchain by pulling on its ends is a standard subject in polymer physics, with important applications in cell biology; as a recent example see, e.g., \cite{stretching} about chromatin stretching by optical tweezers.  By analogy with macroscopic examples (e.g., mooring line around a bollard), here we examine another efficient way to stretch a polymer, when it is tight around a curved obstacle.  As a model, it is prototypical for several biological contexts.  As just one example, we mention the recently documented (via imaging \cite{wong19,oakley70} and Hi-C experiments \cite{nand21}) chromosomes morphology in a certain algae (dinoflagelletes) that form cylindrical rods with helically twisted bundles of wrapped DNA. In addition, the model turns out to have surprisingly rich connections with several other fields of theoretical physics, first and foremost with KPZ scaling.

Winding of a an ideal polymer around a topological point-like obstacle in 2D is also a classical problem, pioneered by S.F. Edwards \cite{Edwards:1967}, Prager and Frisch \cite{Prager_Frisch} and Saito and Chen \cite{Chen}.  For a finite size obstacle in 2D, Green's function of an ideal polymer was considered by Comtet et al in \cite{Comtet} and later formally expressed in terms of an infinite series of linear combinations of Bessel functions \cite{Grosberg_Frish:2003}.  Although the latter result is exact, addressing the limit of strongly stretched chain based on this expression remains a steep challenge. A significant progress in this direction was recently achieved by B.Meerson and N.Smith \cite{Baruch1,Baruch2}, some related problems were also examined by some of us \cite{1,2,valov_fixman}.

\begin{figure}
  \centering
  \includegraphics[width=250pt]{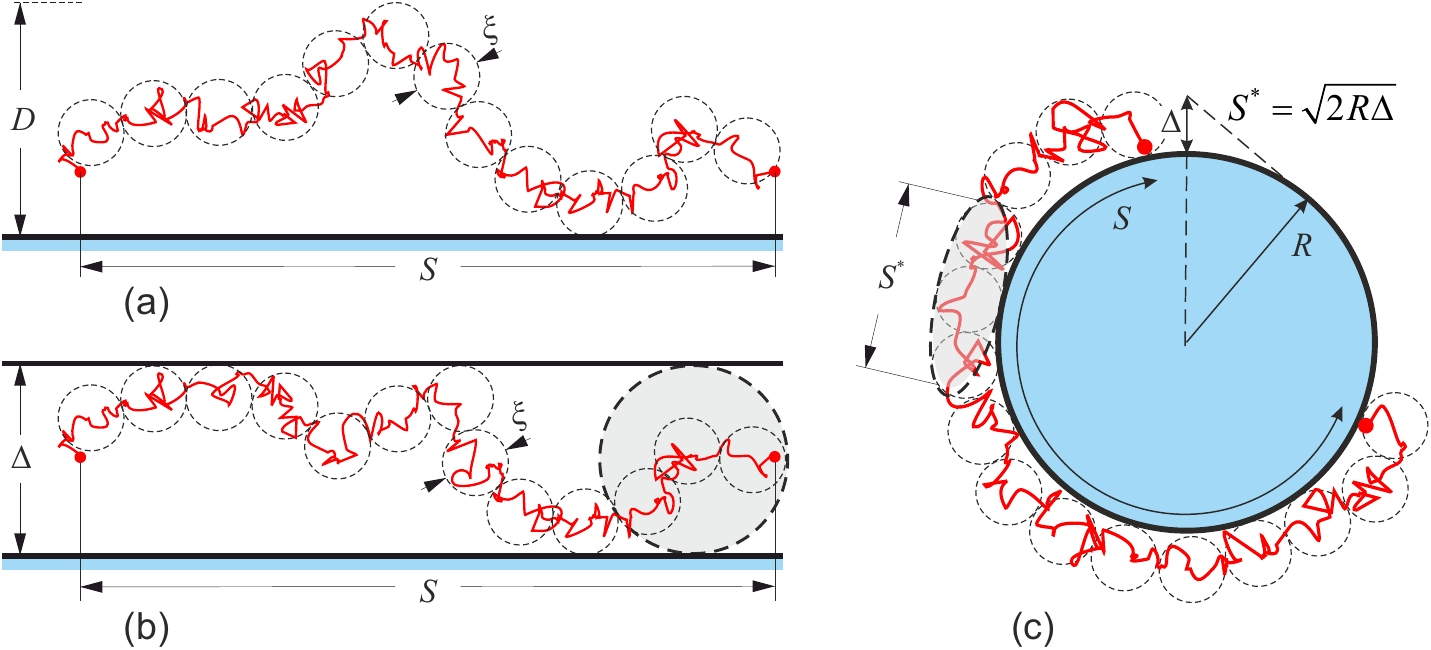}
  \caption{Stretching of a polymer chain in a flat (left) or curved (right) geometry.  In each case, chain is represented as a train of Pincus blobs.  (a): the polymer is stretched above a planar boundary and fluctuates at distance $D$ in the perpendicular direction; (b): the polymer is additionally confined within distance $\Delta \ll D$ above the surface, and Pincus blobs are combined as ``super-blobs'' (grey ball). (c): the polymer is stretched around a circular boundary of radius $R$. End-to-end distance along the surface in all cases is fixed, $S \gg b N^{\nu}$.}
\label{fig:pol-f01}
\end{figure}

The model that we address in this paper is depicted in Fig. \ref{fig:pol-f01}, with panels (\textit{a,b}) and (\textit{c}) showing polymer chain stretched along a flat and curved boundary, respectively.  We will be interested in the span of fluctuations of the polymer away from the boundary, characterized by the length scale $\Delta$.  Following a note by one of us \cite{Grosberg_Comment_2021}, we argue that $\Delta$ in the curved boundary case is determined by the non-local competition between entropy loss when polymer is tightly confined in a narrow strip of width $\Delta$ along the surface, and entropy loss of its stretching beyond imposed necessity by making a wider detour around the obstacle. Our analysis reveals the universal scaling, $\Delta \sim R^{\beta}$, as a function of $R$ with the KPZ growth exponent $\beta=1/3$, while the correlation length $S^*$ at which the chain experiences curvature of the disk scales as $S^* \sim R^{1/z}$ with the KPZ dynamic exponent $z=3/2$. Simulations of this system reveal that the one-point distribution of typical fluctuations can be well described by the squared Airy law, connecting our polymer problem in 2D with the (1+1)D Ferrari-Spohn universality class \cite{spohn_ferrari,Baruch1}.

Strikingly, we found that this KPZ-like behavior is valid not only for an ideal polymer, but for a polymer with an arbitrary fractal dimension $D_f=1/\nu$, where $\nu$ is the usual metric exponent of the mean-square end-to-end distance against the number of monomers $R_0^2 \sim b^2 N^{2\nu}$, where $b$ is the (Kuhn) monomer length. The examples, in addition to the usual ideal $\nu = 1/2$, include self-avoiding polymers in 2D, $\nu = 3/4$ \cite{nienhuis}, in 3D (if the ring is wound around a cylinder), $\nu \approx 0.588$ \cite{li}; annealed branched polymers, $\nu = 1/4$ \cite{zimm} without or $\nu \approx 7/13$ with excluded volume \cite{BranchedUnivClass}; one ring in a 3D melt of other unconcatenated rings, $\nu = 1/3$ \cite{khokhlov-nech,halverson}; polymers with quadratic non-local Hamiltonian producing subdiffusive fractal paths with arbitrary $\nu < 1/2$ \cite{nech-tamm-pol,polovnikov19}.

\textit{Polymer stretching along a flat boundary.}  As a warm-up and for future reference, consider a chain of $M \gg 1$ monomers with the fractal dimension $D_f=1/\nu$ and let it be stretched along a flat boundary, with end-to-end distance fixed at $S \gg b M^{\nu}$. As shown in Fig. \ref{fig:pol-f01}(a), the chain looks like a train of blobs of $g$ monomers of size $\xi$ each. Chain statistics is unaffected inside the blob, i.e. $\xi = b g^{\nu}$, and the train of blobs is stretched, meaning that $S = \frac{M}{g} \xi$. Simple algebra then gives  $g = \left( M b/S \right)^{1/(1-\nu)}$ and $\xi = b \left( M b/S \right)^{\nu/(1-\nu)}$. These blobs generalize the classical ``Pincus blobs'' \cite{PincusBlobs_ma60051a002} for arbitrary $\nu$.

Free energy of chain stretching, $F_{\rm str}$, is about $k_B T$ per blob:
\be
\frac{F_{\rm str}}{k_B T} = \frac{M}{g} = \left( \frac{S}{b M^{\nu}} \right)^{\frac{1}{1-\nu}} \ .
\label{eq:FreeEnergy_Stretch}
\ee
The statistics of chain of blobs in the direction perpendicular to the surface is ideal (compare \cite[Equation 1.50]{deGennes}), its spread is, therefore,
\be
D = \left(\frac{M}{g} \right)^{1/2} \xi = b M^{\nu} \left( \frac{b M^{\nu}}{S} \right)^{\frac{2 \nu - 1}{2 (1-\nu)}} \ .
\label{eq:D_Flat}
\ee
If a chain is additionally confined within a layer of width $\Delta \ll D$ as depicted in Fig. \ref{fig:pol-f01}(b), then, considering ``super-blobs'' of $G$ Pincus blobs each (see Fig. \ref{fig:pol-f01}(b)), such that $\xi G^{1/2} = \Delta$, we can find the confinement free energy as $k_BT$ per super-blob:
\be
\frac{F_{\rm conf}}{k_B T} = \frac{M}{gG} = \frac{b^2 M^{2\nu}}{\Delta^2} \left( \frac{b M^{\nu}}{S} \right)^{\frac{2 \nu -1}{1-\nu}}
\label{eq:FreeEnergy_Conf}
\ee

An interesting observation is that size $D$ perpendicular to stretching (see \eq{eq:D_Flat}) is a \textit{decreasing} function of elongation $S$ only for $\nu > 1/2$, while for $\nu < 1/2$ it is an \textit{increasing} function. In other words, the ``subdiffusive'' polymers with $\nu < 1/2$ behave as a substances with a negative Poisson ratio: upon stretching they swell in perpendicular direction. Clearly, this is because fractal polymers with $\nu < 1/2$ have some negative correlations between monomers. These correlations are destroyed by stretching which leads to chain's ``releasing''.

\begin{figure*}
  \centering
  \includegraphics[width=\textwidth]{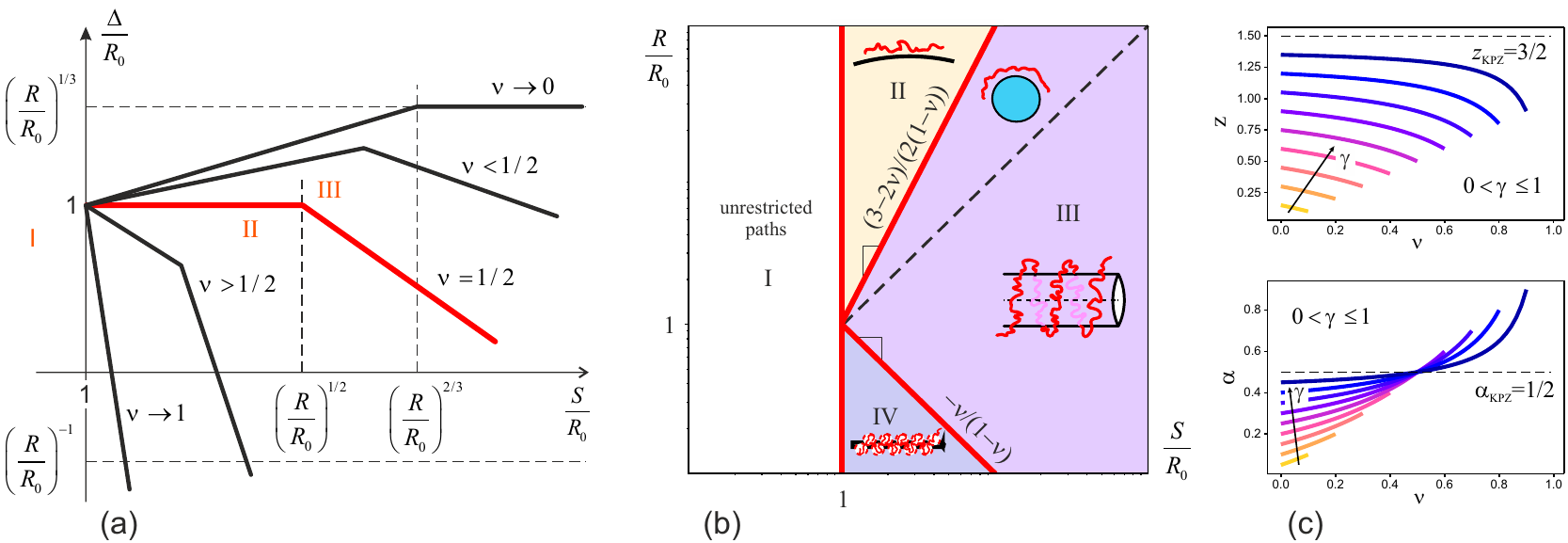}
  \caption{\textbf{(a)}: Polymer chain spread, $\Delta$, away from a cylindrical surface as a function of curvilinear distance between chain ends $S$, for a variety of $\nu$ values.  All distances measured relative to the unperturbed coil size $R_0 = b N^{\nu}$. Three main regimes of \eq{eq:sumDelta} are shown for $\nu=1/2$ in red.  \textbf{(b)}: Four regimes of fluctuations for various values of disc radius $R$ and end-to-end distance $S$. Regime II corresponds to effectively flat barrier, while regime IV is for an obstacle thinner than one Pincus blob.  The most interesting is regime III, it corresponds to the stretched polymer on the essentially curved barrier. The dashed line highlights winding around the cylinder. (c): Exponents $z$ (upper) and $\alpha$ (bottom) as the functions of $\nu <\gamma$ for different values of $\gamma$. The increase of $\gamma$ from $\gamma=0.1$ (yellow) to $\gamma=0.9$ (dark blue) is shown by arrows on both diagrams. Limiting KPZ values corresponding to $\gamma=1$ are marked by black dashed lines.}
  \label{fig:pol-f02}
\end{figure*}


\textit{Polymer stretching along a curved boundary.} Let us keep ends of a chain of $N$ monomers at distance $S$ away along a surface of a cylinder of radius $R$ -- see Fig. \ref{fig:pol-f01}(c). We assume the chain is stretched, i.e. $S \gg a N^{\nu}$, however we do not impose any relation between $S$ and $R$. Free energy of such a chain consists of two contributions. The first one describes chain stretching to the distance $\frac{S}{R} \left(R+\Delta \right)$; the corresponding free energy is given by \eq{eq:FreeEnergy_Stretch}, by replacing $M \to N$ and $S \to S \left(1 + \Delta/R \right)$. The second term corresponds the polymer confining in a strip of the width $\Delta$, and it is given by \eq{eq:FreeEnergy_Conf}, with similar replacement $M \to N$. Overall, variational free energy becomes
\be
\frac{F_{\mathrm{circ}}}{k_B T} =  \left(\frac{S}{b N^{\nu}}\; \frac{R+\Delta}{R} \right)^{\frac{1}{1-\nu}}  + \frac{b^2 N^{2\nu}}{\Delta^2} \left( \frac{b N^{\nu}}{S} \right)^{\frac{2 \nu -1}{1-\nu}}
\label{eq:free_energy_circle}
\ee
This free energy is the extension of Eq. (2) in the Comment \cite{Grosberg_Comment_2021}. Assuming $\Delta \ll R$, we can linearize the first term and then minimize this free energy to get
\be
\frac{\Delta}{bN^{\nu}} = \left( \frac{R}{bN^{\nu}} \right)^{\frac{1}{3}} \left( \frac{b N^{\nu}}{S} \right)^{\frac{2 \nu}{3(1-\nu)}}
\label{eq:Delta_Curved}
\ee

It is instructive to re-derive (\ref{eq:Delta_Curved}) in a different way. Given the chain is localized in a strip of width $\Delta$, curvature of the underlying surface becomes relevant only at scales exceeding $S^{\ast} = (R \Delta )^{1/2}$ (see Fig. \ref{fig:pol-f01}(c)).  We call chain section covering distance $S^{\ast}$ ``an elliptic blob''.  To find the number of monomers in elliptic blob,  $N^{\ast}$, we can use the result \eq{eq:D_Flat}, with the replacement $D \to \Delta$ and $S \to S^{\ast}$.  The train of elliptic blobs is fully stretched around the curved boundary, hence their number is $N/N^{\ast} = S/S^{\ast}$.  A few lines of algebra based on this relation yield the previously obtained answer (\ref{eq:Delta_Curved}). Simultaneously, we get the expressions for cross-over scale $S^{\ast}$ at the given $R$:
\be
S^{\ast}/b N^{\nu} = \left( R / b N^{\nu} \right)^{\frac{2(1-\nu)}{3-2\nu}}\ .
\label{eq:crossover}
\ee

Our results (\ref{eq:D_Flat}) for effectively flat surface at $S < S^{\ast}$ and (\ref{eq:Delta_Curved}) for the curved one at $S > S^{\ast}$ are collected in Fig. \ref{fig:pol-f02}(\textit{a}).  There, we plot $\Delta /bN^{\nu}$ as a function of stretching $S/bN^{\nu} > 1$ for a variety of $\nu$ values, $\nu\in (0,1)$.  In particular, at $\nu < 1/2$, the behavior $\Delta(S)$ is non-monotonic: $\Delta$ increases at modest $S$, because stretching destroys negative correlations, while at larger $S$ the curvature of the underlying disc takes over and forces $\Delta$ to decrease again. Thus, for ``subdiffusive'' paths the negative Poisson ratio flips its sign to the positive at stretching $S^*$ when the boundary becomes substantially curved. A curious fact is that at the specific value  $\nu = 1/3$ (which corresponds to a sort of a crumpled statistics) the non-monotonous dependence $\Delta(S)$ comes back to the starting value $\Delta / b N^{\nu} \sim 1$ exactly when $S$ becomes of the order of disc radius, $R$, i.e. when chain is forced to make about one full turn around the disc.

Another way to summarize our results is given in the  Fig. \ref{fig:pol-f02}(\textit{b}) in terms of four distinct regimes in the space of two dimensionless control parameters, the amount of stretching $S$ and radius of the disc $R$, both scaled by the unperturbed coil size, $S/bN^{\nu}$ and $R/bN^{\nu}$:
\be
\frac{\Delta}{b N^{\nu}} = \left\{\begin{array}{ll}  1 , & \mathrm{Regime \ I} \\\left( S / b N^{\nu} \right)^{-\frac{2 \nu - 1}{2 (1-\nu)}}, & \mathrm{Regime \ II}  \\   \left(R / b N^{\nu} \right)^{\frac{1}{3}} \left( S / b N^{\nu} \right)^{-\frac{2 \nu}{3 (1-\nu)}} , & \mathrm{Regime \ III} \\ \left( S / b N^{\nu} \right)^{-\frac{\nu}{1-\nu} }, & \mathrm{Regime \ IV}  \end{array} \right.
\label{eq:sumDelta}
\ee

The first regime (I) deals with the unrestricted polymers with $S< b N^{\nu}$; for them, fixation of ends only marginally affects the statistics, $\Delta \sim b N^{\nu}$. The second (II) and the third (III) regimes correspond to stretched polymers above effectively flat \eq{eq:D_Flat} and curved \eq{eq:Delta_Curved} boundaries, respectively. The most interesting regime, where the span of fluctuations is $R$-dependent, is the regime (III); remarkably, in this regime the dependence $\Delta(R) \propto R^{1/3}$ turns out to be independent on $\nu$.

When the cylinder radius becomes as small as the Pincus blob size, $R \sim \xi$, so does the elliptic blob, $S^{\ast} \sim \xi$, and the crossover to regime IV occurs. Clearly, in this regime every Pincus blob ``hugs'' around the entire cylinder and, thus, $\Delta = \xi \ge R$, \footnote{Note that \eq{eq:Delta_Curved} is not valid in this regime, since the Pincus blob $\xi$ exceeds the size of the elliptic blob $S^{\ast}$.}. In terms of the winding number, $n$, the regime IV corresponds to $n > \left(S/bN^\nu\right)^{1/(1-\nu)} \gg 1$. In this regard, it is tempting to compare regime IV of winding around a thin cylinder with winding around a zero-width line studied earlier \cite{Edwards:1967, Prager_Frisch, Chen, Comtet, Grosberg_Frish:2003}. In that case, $\Delta \sim b N^{\nu}$, with only weak dependence on winding number $n$ (see formula (48) in the work \cite{Grosberg_Frish:2003}, which is exact for $\nu = 1/2$). Since physical obstacle is always not purely topological, but also geometrical with some finite radius $R$, our present work allows to clarify the applicability conditions of the result of \cite{Grosberg_Frish:2003}: $2 \pi R n \ll b N^{\nu}$.

\begin{figure}
  \centering
  \includegraphics[width=250pt]{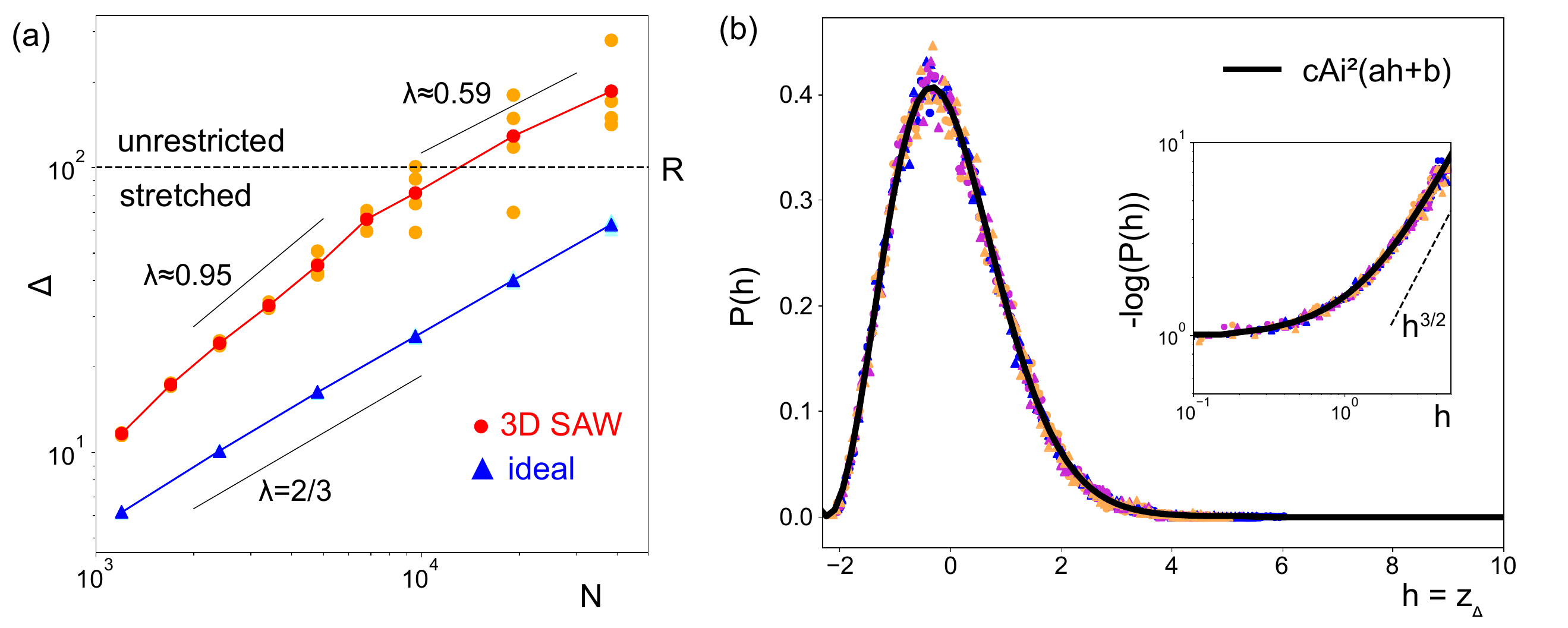}
  \caption{Molecular dynamics simulations of a stretched polymer over a disc of radius $R = 100b$ with two ends fixed at diametrically opposite points. \textbf{(a)}: Average excursions as a function of $N$ for $\nu=1/2$ (triangles) and $\nu\approx 0.588$ (circles). Cyan and orange symbols show results of independent runs; blue and red symbols are the averages. According to \eq{eq:Delta_Curved}, our theory predicts at fixed $R$ and $S = \pi R$ that $\Delta \sim N^{\lambda} b$, with $\lambda = 2 \nu/3(1-\nu)$ for stretched chains ($\Delta < R$) and $\lambda = \nu$ for unrestricted ones ($\Delta > R$). \textbf{(b)}: Probability density $P(h)$ of standardized (z-score) excursions $h=z_\Delta$ of the median monomer above the disk. The collapse is shown for polymer lengths $N=1200,2400,4800$ (each coded by individual color), and two values of $\nu$: $\nu=0.5$ (ideal chain, triangles) and $\nu\approx 0.588$ (3D self-avoiding walk, circles). The black thick line is the properly scaled squared Airy function with the parameters $a \approx 0.7$, $b \approx 0.78$, $c \approx 1.42$ (see \cite[Section S4]{SupMat}). The inset shows the tail of $-\log P(h)$, which asymptotically goes as $h^{3/2}$.}
  \label{fig:pol-f03}
\end{figure}

Our results can be generalized for the chain stretched around a smooth barrier of varying curvature, e.g., around an ellipse, see Appendix \ref{app_curved} and Fig. A1.

Although tightening polymer chain around a cylinder forces it to lean on the surface, its conformation is very different from a chain adsorbed on the surface by short range attractive potential.  As we show in Appendix \ref{app_adsorbed}, end stretching acts not as a short range, but rather a scale-free potential.  This is consistent with the fact that all our results above, as in the classical Pincus theory, are derived from extensive (linear in $N$) terms in free energy, not involving critical exponents related to $\gamma$ (reviewed e.g., in \cite{Paper_about_gamma}).

In order to illustrate the analytical scaling results we have performed molecular dynamics simulations (see Appendix \ref{app_sims} for details) of stretched ideal polymers ($\nu = 1/2$) and 3D self-avoiding chains ($\nu \approx 0.588$). Namely, we fixed chain ends at two diametrically opposite points of a disc, $S=\pi R$, at fixed $R$, and examined the $N$ dependence of the averaged height $\Delta$ for the middle monomer.  As one can see from Fig. \ref{fig:pol-f03}(\textit{a}) and Fig. A2 in Appendix the span of fluctuations in both sets of simulations is consistent with the predictions \eq{eq:Delta_Curved}. As an additional benefit of having performed the simulation, we further compute the distribution of the excursions at various $N$, see Fig. \ref{fig:pol-f03}(\textit{b}) and found them close to squared Airy function, consistent with the results of Ferrary and Spohn \cite{spohn_ferrari} as well as subsequent works \cite{valov_fixman,Baruch1}.  While all of these earlier works considered only the Brownian case $\nu = 1/2$, our numerical results include also $\nu = 0.588$ and thus hint on the possibility that squared-Airy distribution, and not only scaling exponents, may be universal across polymer fractalities with different $\nu$ values (see further details in Appendix \ref{app_airy}).


\textit{Discussion.} One of the most striking of our results is that the dependence of the typical fluctuations in the curved regime on the disc radius, $R$, can be written as $\Delta = R^{\beta}\, f(S,N,b,\nu)$ with $\beta=1/3$ being the 1D KPZ \textit{growth} exponent. Thus, it is tempting to look for a mapping between our problem and KPZ and to interpret $R$ as time, $t$, in the associated stochastic growth problem. To see how it works, let us set $S = N^{\gamma} b \gg N^{\nu} b$, i.e. consider $\nu < \gamma < 1$. As \eq{eq:Delta_Curved} suggests, in the limit $\gamma \to 1$ typical fluctuations $\Delta$ are controlled by $R$ only
\be
\Delta = b(R/b)^{\beta},
\ee
for $R < R^*$, where $R^* = b(S/b)^z$ is the crossover radius, below which a polymer with $N=(S/b)^{1/\gamma}$ monomers experiences curvature of the disk, and $z$ reads
\be
z(\gamma, \nu)= \frac{3\gamma-2\nu \gamma-\nu}{2\gamma(1-\nu)} \to 3/2, \quad \gamma \to 1.
\label{rstar}
\ee
The crossover, described by the 1D KPZ \textit{dynamic} exponent, $z=3/2$, \eq{rstar} corresponds to the boundary between flat (II) and curved (III) regimes in \fig{fig:pol-f02}(\textit{b}). In the flat regime (II), $R>R^{\ast}$, the typical fluctuations do not depend on $R$ and are described by the stretching $S$ only, $\Delta = b(S/b)^{\alpha}$, where $\alpha$ reads
\be
\alpha(\gamma, \nu) = 1 - \frac{\gamma-\nu}{2\gamma(1-\nu)} \to \frac{1}{2}, \quad \gamma \to 1,
\ee
yielding the 1D KPZ \textit{roughness} exponent $\alpha=1/2$. In Fig. \ref{fig:pol-f02}(c) we demonstrate the dependence of the exponents $z$ and $\alpha$ on $\nu$ at different values of $\gamma$. As an intrinsic property of the ``full stretching'' limit, the curves $z(\nu), \alpha(\nu)$ become flat upon the increase of $\gamma$, approaching their respective 1D KPZ values.

Importantly, the implications of the $\gamma \to 1$ limit above can be realized for any other $\gamma$, but the chain should be renormalized to the Pincus blobs. Indeed, under the change $N \to N/g$, $b \to \xi$ a two-dimensional walk becomes effectively (1+1)D and, therefore, it naturally inherits all the scalings of the ``full stretching'' limit
\be
\begin{cases}
\Delta_{\mathrm{curved}} = \xi (R/\xi)^{\beta}, \quad S>S^* \medskip \\
S^* = \xi (R/\xi)^{1/z} \medskip \\
\Delta_{\mathrm{flat}} = \xi (S/\xi)^{\alpha}, \quad S<S^*
\end{cases}
\label{eq:renorm}
\ee
where $\xi = \xi(N,S)$ plays a role of a new coarse-grained monomer.  Note that \eq{eq:renorm} holds for any fractal dimension of the polymer.

From representation \eq{eq:renorm} it is evident that upon proper renormalization \textit{any} fractal walk in two dimensions above the disk can be described by the set of KPZ exponents.  However, as our simulations suggest, the distribution of typical fluctuations in the polymer problem is given by the squared Airy function, which is different from the Tracy-Widom distribution of the KPZ process (though the tails, $\sim \exp(-ch^{3/2})$, are equivalent, see the inset in \ref{fig:pol-f03}(\textit{b})). In fact, this is a well-known consequence of the impermeability of the boundary, playing a role of the ``mean-field'' for a more complex system of many non-intersecting (``vicious'') (1+1)D Brownian walks, the top of which is known to belong to the KPZ universality class (see the flowchart and further discussion in Appendix \ref{app_bv} and \ref{app_flow}). Replacing all such walks (but the top one) with the circular boundary we arrive at the Ferrari-Spohn model, which demonstrates the squared Airy behaviour for the one-point distribution \cite{spohn_ferrari,Baruch2}. Therefore, though our model does not belong to the KPZ class, we conjecture that it shares the Ferrari-Spohn universality.

Another interesting connection of our problem is revealed by looking at free energy (\ref{eq:free_energy_circle}) for the specific case when $\nu = 1/2$ and radius has specific value $R = S^2/b N$ (indicated by a dashed blue line on \fig{fig:diagram_bv}).  Along this line, free energy reads $\frac{F_{\mathrm{circ}}}{k_BT} = \frac{\Delta}{b} + \frac{b^2 N}{\Delta^2}$, which can be interpreted by noticing that $W(N) = \max_{\Delta} \exp \left( -\frac{\Delta}{b} - \frac{b^2 N}{\Delta^2} \right)$ is the probability for a random walker with diffusivity $b^2/\pi^2$ to survive during time $N$ on the line with randomly Poisson positioned traps with density $1/b^2$.  This is classical Balagurov-Vaks problem \cite{balagurov}, and its solution $W(N) \sim e^{ - \mathrm{const} \, N^{1/3}}$ is controlled by the optimal interval between the traps, $\Delta$.  In Appendix \ref{app_bv} and \ref{app_flow} we develop this connection and review several relations to other problems and models in statistical physics.

\begin{acknowledgements}


The authors thank A. Gorsky, M. Tamm and A. Valov for illuminating discussions. The work of KP is supported by the Russian Science Foundation (Grant No. 21-73-00176). AYG acknowledges the Aspen Center for Physics where part of this work was written with the support of the National Science Foundation grant number PHY-1607611. The authors thank MirnyLab for kindly sharing the resources for computer simulations.

\end{acknowledgements}

\renewcommand{\theequation}{A-\arabic{equation}}
\renewcommand{\thefigure}{A\arabic{figure}}
\renewcommand{\thesection}{A\arabic{section}}

\setcounter{equation}{0}  
\setcounter{figure}{0}  
\setcounter{section}{0}


\section*{APPENDIX}  


In this Appendix we provide some additional information related to our paper, as follows:  \begin{enumerate} \item We generalize our results about stretching polymer around a circular disc and consider stretching it around an ellipse or another convex barrier.  \item We discuss the analogy and fundamental difference between chain tightened around an obstacle and chain adsorbed on the obstacle by a short range potential.  \item  We describe technical details of polymer simulations used in our work.  \item We summarize the statistical details related to numerical analysis of probability distributions of polymer excursions away from the curved surface.  \item We re-formulate our problem and establish its connection to the problems of random walks in the space with Poisson-distributed traps.  \item Finally, we speculate about other far-reaching connections of our problem across the fields.  \end{enumerate}
 
\section{Stretching of a polymer around an elliptic (or another convex) barrier}\label{app_curved}

Let us return to Eq. (4) in the main text. We offer here a slightly different view on it. Let us start from a scaling derivation of Pincus blobs. Clearly, the quantity $\xi$ is a correlation length. Given that there is only one macroscopic length scale for an unrestricted coil, $R_F = b N^{\nu}$, correlation length in case when two ends stay at distance $S$ apart, must obey the scaling $\xi = R_F \phi (S/R_F)$, where the behavior of scaling function $\phi(x)$ is as follows: $\phi(x) \sim 1$ when $x \ll 1$, while $\phi(x)\sim x^{\mu}$ when $x \gg 1$ with some critical exponent $\mu$. The latter must be chosen such that for $S \to b N$ the blob size is reduced to (Kuhn) monomer size $b$. The following equation with $\mu = \nu / (1-\nu)$, provides the requested behavior:
\be
\xi = R_F \left( \frac{R_F}{S} \right)^{\frac{\nu}{1-\nu}}
\label{eq:xi}
\ee
If the chain is stretched by force rather than by fixing end-to-end distance, then $\xi = k_B T/f$, so $\xi$ can be viewed as a proxy for the stretching force.

Now, we return to the equation (4) of the main text and re-write it as follows:
\be \begin{split}
\frac{F_{\mathrm{circ}}}{k_B T} & = \left(\frac{S}{b N^{\nu}}\; \frac{R+\Delta}{R} \right)^{\frac{1}{1-\nu}} + \frac{b^2 N^{2\nu}}{\Delta^2} \left(\frac{b N^{\nu}}{S} \right)^{\frac{2 \nu -1}{1-\nu}}  \\ &  = \left( \frac{S}{b N^{\nu}} \right)^{\frac{1}{1-\nu}} \frac{\Delta}{R} + \frac{b^2 N^{2\nu}}{\Delta^2} \left(\frac{b N^{\nu}}{S} \right)^{\frac{2\nu-1}{1-\nu}}  \\ & = \frac{S}{\xi} \left(\frac{\Delta}{R} + \frac{\xi^2}{\Delta^2}\right) \ .
\label{eq:free_energy_circle_1}
\end{split}
\ee
In the last transformation, we expressed free energy in terms of blob size, $\xi$.  Minimization of this free energy with respect to $\Delta$ yields the result
\be
\Delta = \left(R \xi^2 \right)^{1/3}
\label{eq:xi2}
\ee
Of course, this result is entirely equivalent to our previous formula (5) in the main text for the curved surface. Quite remarkable fact is that this result does not involve $\nu$ at all. In terms of $\xi$ (or stretching force) there is no dependence on $\nu$.

\begin{figure}[ht]
  \centering
  \includegraphics[width=0.4\textwidth]{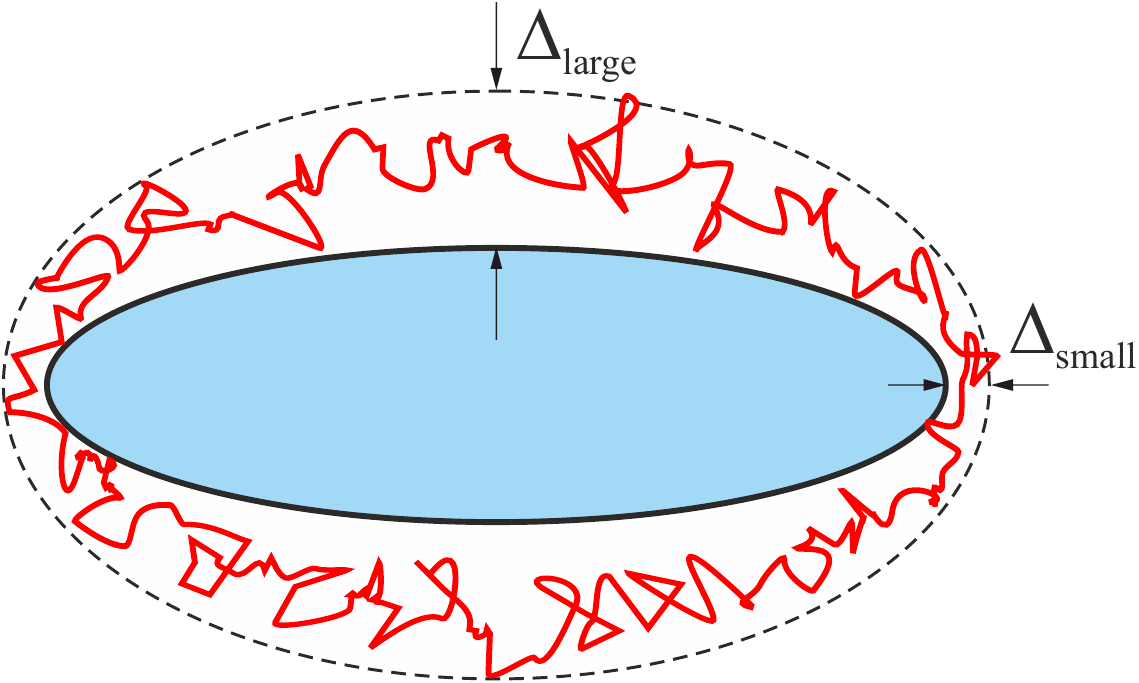}
  \caption{Polymer chain stretched around an ellipse.}
  \label{fig:Bumpy_Surface}
\end{figure}

The result \eq{eq:xi2} allows us to consider stretching a polymer around a bumpy surface whose curvature changes from place to place, for instance, around an ellipse or around other convex curve with slowly changing curvature as shown in \fig{fig:Bumpy_Surface}.

Since tension force is the same everywhere along the polymer, so is the blob size $\xi$. Therefore, if curvature radius is different in different places (while changing slowly, over length scales much larger than $\xi$), the factor $S/\xi$ in \eq{eq:free_energy_circle_1} can be replaced by the integral along $S$ with ``density'' $1/\xi$:
\be
\frac{F}{k_B T}  =  \int_{0}^{S} \frac{dx}{\xi} \left[ \frac{\Delta(x)}{R(x)}  +  \frac{\xi^2 }{\Delta^2(x)}\right] \ , \label{eq:free_energy_circle_2}
\ee
providing the result
\be
\Delta(x) = \left(R(x) \xi^2 \right)^{1/3} \ .
\label{eq:curv}
\ee
These results apply of course only to the case of everywhere convex impermeable boundary, because if some parts are concave, the stretched polymer will take a straight shortcut.

\section{Is tightly wound chain similar to the adsorbed one?}\label{app_adsorbed}

It is intuitively appealing to think of the chain tightly stretched around a cylinder as similar to a chain adsorbed on the cylinder surface.  In fact, fundamental difference between these two problems is that adsorption is usually due to a short range potential attracting monomers to the surface, while stretching by the ends, if anything, is similar to a scale free long range potential.  In this section, we briefly explain this analogy for the simplest case of a strongly stretched Gaussian polymer.

Let us describe the impermeable cylindrical obstacle as a potential barrier
\begin{equation} U_0(r) = \left\{ \begin{array}{lcr} + \infty & \text{at} & r<R \\ 0 & \text{at} & r > R \end{array} \right. \end{equation}
which prohibits monomers from entering the $r<R$ region.  Following Edwards \cite{DoiEdwards}, partition sum of Gaussian chain can be written as a path integral
\begin{equation} Z = \int D \mathbf{r}(s) \exp \left[- \int_{0}^{N} \left( \frac{3 \dot{\mathbf{r}}^2(s)}{2 b^2} + U_0(\mathbf{r}(s)) \right) ds \right] \label{eq:general_path_integral} \end{equation}
Here $s$ is the monomer number along the chain, and dot indicates derivative with respect to $s$.   In polar coordinates $\mathbf{r} \to (r,\theta)$, we have $r(s) = R + \Delta(s)$, and so
\begin{equation}\begin{split}  \dot{\mathbf{r}}^2(s) & = \dot{\Delta}^2(s) + (R+ \Delta(s))^2 \dot{\theta}^2(s) \\ & \simeq R^2 \dot{\theta}^2(s) + \dot{\Delta}^2(s) + 2 R \Delta(s) \dot{\theta}^2(s) \end{split}
\label{eq:kin_energy_polar_coord} \end{equation}
When we plug this in the path integral (\ref{eq:general_path_integral}), we have to integrate over both $\theta(s)$ and $\Delta(s)$, and these variables are coupled.  The situation is simplified in the strong stretching regime, when azimuthal fluctuations are suppressed, and $\theta(s) \simeq sb/R$.  In this approximation, partition sum (\ref{eq:general_path_integral}) (ignoring constant factors) becomes
\begin{equation}  Z  =  \int D \Delta(s) \exp \left[- \int_{0}^{N} \left( \frac{3 \dot{\Delta}^2(s)}{2 b^2} + U_{\mathrm{eff}}(\Delta(s)) \right) ds \right] \ , \label{eq:effective_path_integral}  \end{equation}
where
\begin{equation} U_{\mathrm{eff}} (\Delta) = \left\{ \begin{array}{lcr} + \infty & \text{at} & \Delta < 0  \\ 3 \Delta / R   & \text{at} & \Delta > 0 \end{array} \right. \label{eq:triangular_potential} \end{equation}
Partition sum (\ref{eq:effective_path_integral}) describes a Gaussian polymer in one dimension confined in a triangular potential (\ref{eq:triangular_potential}).  This potential obviously holds a discrete spectrum for the Schr\"{o}dinger equation
\begin{equation} \frac{b^2}{6} \frac{d^2 \psi}{d \Delta^2} - U_{\mathrm{eff}} (\Delta) \psi = -\lambda \psi \ , \label{eq:Schrodinger} \end{equation}
and its ground state eigenvalue gives the relevant free energy.  Note that scaling estimates of the terms of this equation ($b^2/\Delta^2$ for the first term, and $\Delta / R$ for the second) are exactly the same as scaling estimates of free energy above in this work (e.g., formula (\ref{eq:free_energy_circle_1})).

As a side note, the ground state eigenfunction of Schr\"{o}dinger equation (\ref{eq:Schrodinger}) in the triangular potential is the Airy function ($\psi(\Delta) = \mathrm{Ai} \left( \left( \frac{18}{b^2 R} \right)^{1/3} \Delta \right)$), while probability density $P(\Delta)$ is given by $\psi^2(\Delta)$ (see, e.g., \cite{RedBook}), i.e., by the squared Airy function.  This result is not surprising, because strongly stretched polymer, with suppressed azimuthal fluctuations, is similar to the (1+1)D random path, for which the squared Airy distribution was established by Ferrari and Spohn \cite{spohn_ferrari}.

To summarize, stretching of the chain around a cylinder is equivalent to confining polymer at the surface by effective potential (\ref{eq:triangular_potential}) that is not localized at the surface.   We showed it for the limit of very strong stretching.  To relax it, one should realize that $\theta(s)$ fluctuates over the scale of the order of Pincus blob, and becomes a linear function of $s$ only at larger scales.  We do not delve into the analysis of this situation, because our only goal here is to illustrate that chain tightening around a cylinder is associated with a scale-free effective potential, and this fact is clear from the consideration of the simplest case.

As a reminder, when adsorbing potential is localized at the surface, chain conformation is usually considered in terms of trains (localized inside the potential well) and loops (freely fluctuating outside of potential well).  This view is equally applicable to the ideal Gaussian chains as well as self-avoiding ones.  Trains and loops are described theoretically in terms of critical exponents $\gamma_{1}$ and $\gamma_{11}$ (see, e.g., \cite{Paper_about_gamma}).  The above consideration shows that our problem cannot be considered in terms of trains and loops, because potential (\ref{eq:triangular_potential}) has no scale.

\section{Details of polymer simulations}\label{app_sims}

Simulations of stretched trajectories are done using polychrom module (available at https://github.com/open2c/polychrom), a wrapper around the open source GPU-assisted molecular dynamics package OpenMM \cite{eastman10}. A chain with phantom beads in simulations is supposed to model the ideal Gaussian chain with the fractal dimension $D_f=2$. The chain is equipped with harmonic bonds of the following energy
\be
U_{\mathrm{bond}} = \frac{3}{2a^2} \sum_{i=1}^{N-1} \left(r_{i,i+1} - l_b\right)^2
\ee
where $a=0.06$ is the standard deviation of the monomer-to-monomer distance, $r_{i,i+1}=|\mathbf{r}_{i+1}-\mathbf{r}_{i}|$; the equilibrium bond length is $l_b=1$.

The cylindrical barrier for the chain is aligned along the $z$-axis, having the infinite length and the radius $R$ in the $x-y$ plane. In order to prohibit the chain entering the area constrained by the cylinder, the following soft repulsive potential of strength $k_{\mathrm{cyl}}=5$ is introduced when the chain crosses the disk boundary
\be
U_{\mathrm{cyl}} = k_{\mathrm{cyl}} \sum_{i=1}^{N} {\cal{H}}\left[R - \sqrt{x_i^2+y_i^2}\right] \left(R - \sqrt{x_i^2+y_i^2}\right)^2
\ee
with ${\cal{H}}[.]$ being the Heaviside step function. This potential has been further smoothed in simulations close to the vicinity of the boundary by means of a small parameter inserted under the root. Also, in order to keep the chain ends at the distance $S=\pi R$ apart, we additionally tether the end beads $\mathbf{r}_1, \mathbf{r}_{N}$ at two points on the diameter by springs of strength $k_{\mathrm{th}}=100$ at a small distance $\delta=0.1 < \Delta$ from the disk surface.

In case of the chains with excluded volume, additional pairwise repulsive force is added. The excluded volume potential $U_{\mathrm{ev}}$ is introduced via the auxiliary Weeks-Chandler-Anderson (WCA) potential $U(r_i, r_j)$ \cite{weeks71,kremer90}, which is a lifted Lennard-Jones repulsive branch
\begin{equation}
\begin{split}
&U(r_{ij}=|\mathbf{r_{i}}-\mathbf{r_{j}}|) = \\
&\begin{cases}
    4 \varepsilon\left((\sigma/r_{ij})^{12} - (\sigma/r_{ij})^6 \right)+\varepsilon, \quad & \text{$r_{ij}\leq 2^{1/6}\sigma$} \\
    0,                 & \text{$r_{ij} > 2^{1/6} \sigma$}
\end{cases}
\end{split}
\label{upair}
\end{equation}
where $\sigma$ is the characteristic scale of the excluded volume repulsion and $\varepsilon=1$. In order to avoid strong repulsive forces close to the singularity of \eq{upair} in simulations, the WCA potential is further smoothly truncated
\begin{equation}
\begin{split}
&U_{\mathrm{ev}}(r_{ij}) =\mathcal{H}(\varepsilon_{\mathrm{tr}}-U(r_{ij}))U(r_{ij}) + \\ & + \mathcal{H}(U(r_{ij})-\varepsilon_{\mathrm{tr}}) \varepsilon_{\mathrm{tr}} \left(1 + \tanh \left[\frac{U(r_{ij})}{\varepsilon_{\mathrm{tr}}}-1\right]\right)  \ ,
\end{split}
\label{hev1}
\end{equation}
at the prescribed truncation value $\varepsilon_{\mathrm{tr}}=5$, corresponding to a strong mutual volume exclusion of the beads; as before, $\mathcal{H}$ is the step function. The potential \eq{hev1} acts between every pair of beads, except for neighboring ones.

The chain of length $N$ is initialized with a random walk configuration and equilibrated for a Rouse time $\tau_R$ in the potentials above. A special attention is paid to the absence of wrapping of a polymer around the cylinder in the initial configuration.

\section{Universal squared Airy PDF}\label{app_airy}

\begin{figure}
\includegraphics[width=0.4\textwidth]{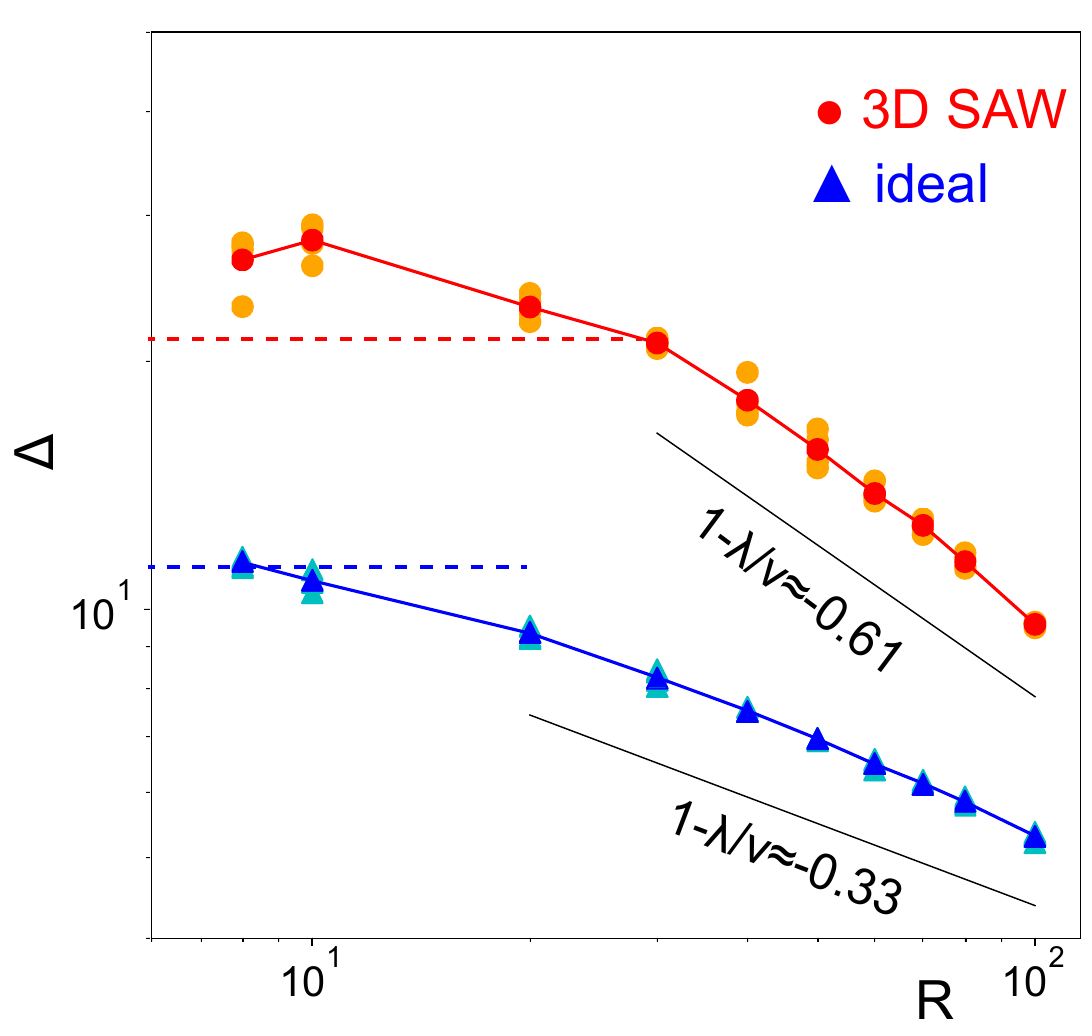}
  \caption{Average excursions as a function of $R$ at the fixed chain length, $N=1000$ and two ends fixed at diametrically opposite points, for $\nu=1/2$ (triangles) and $\nu\approx 0.588$ (circles). Cyan and orange symbols show results of independent runs performed for each $R$ and $\nu$; blue and red symbols are the averages. The theoretical slopes of the dependencies are given by $1-\lambda/\nu$, as dictated by the "boundary condition": at $R\sim b N^\nu$ the excursions should become unconstrained, $\Delta \sim N^{\nu}$.}
  \label{fig:distrs}
\end{figure}

In Fig.3(b) of the main text we collapse distributions of fluctuations of the top monomer $\Delta$ for different values of $N$ by means of standardizing, $\Delta \to h=(\Delta-\langle \Delta \rangle)/\sqrt{\langle (\Delta-\langle \Delta \rangle)^2\rangle}$. We find that the standardized densities collapse onto the universal squared Airy probability density function $c \mathrm{Ai}^2(ah+b)$ with zero mean and unitary variance. The constants $a,b,c$ are determined by the following equations
\begin{equation}
\begin{cases}
\int_{h_0}^{+\infty} h \; \mathrm{Ai}^2(ah+b)\; dh = 0 \\
c \int_{h_0}^{+\infty} h^2 \; \mathrm{Ai}^2(ah+b)\; dh = 1 \\
\int_{h_0}^{+\infty} \mathrm{Ai}^2(ah+b)\; dh = 1/c \\
ah_0+b = \alpha_1,
\end{cases}
\end{equation}
where $\alpha_1 \approx -2.338$ is the first zero of the Airy function.

Solution to this set of equations provides one the values of the constants $a,b,c$
\begin{equation}
\begin{cases}
b = \frac{\int_{\alpha_1}^{+\infty} y \mathrm{Ai}^2(y) dy}{\int_{\alpha_1}^{+\infty} \mathrm{Ai}^2(y) dy} \approx -0.7794 \\
a = \sqrt{\frac{\int_{\alpha_1}^{+\infty} (y-b)^2 \mathrm{Ai}^2(y) dy}{\int_{\alpha_1}^{+\infty} \mathrm{Ai}^2(y) dy}} \approx 0.6971 \\
c = \frac{a}{\int_{\alpha_1}^{+\infty} \mathrm{Ai}^2(y) dy} \approx 1.4177 \\
h_0 = \frac{\alpha_1-b}{a} \approx -2.2358.
\end{cases}
\label{abc}
\end{equation}
We note that the fact that $h_0$ turns out to be rather close to $\alpha_1$ is a coincidence that should not confuse.

\begin{figure*}
\includegraphics[width=0.9\textwidth]{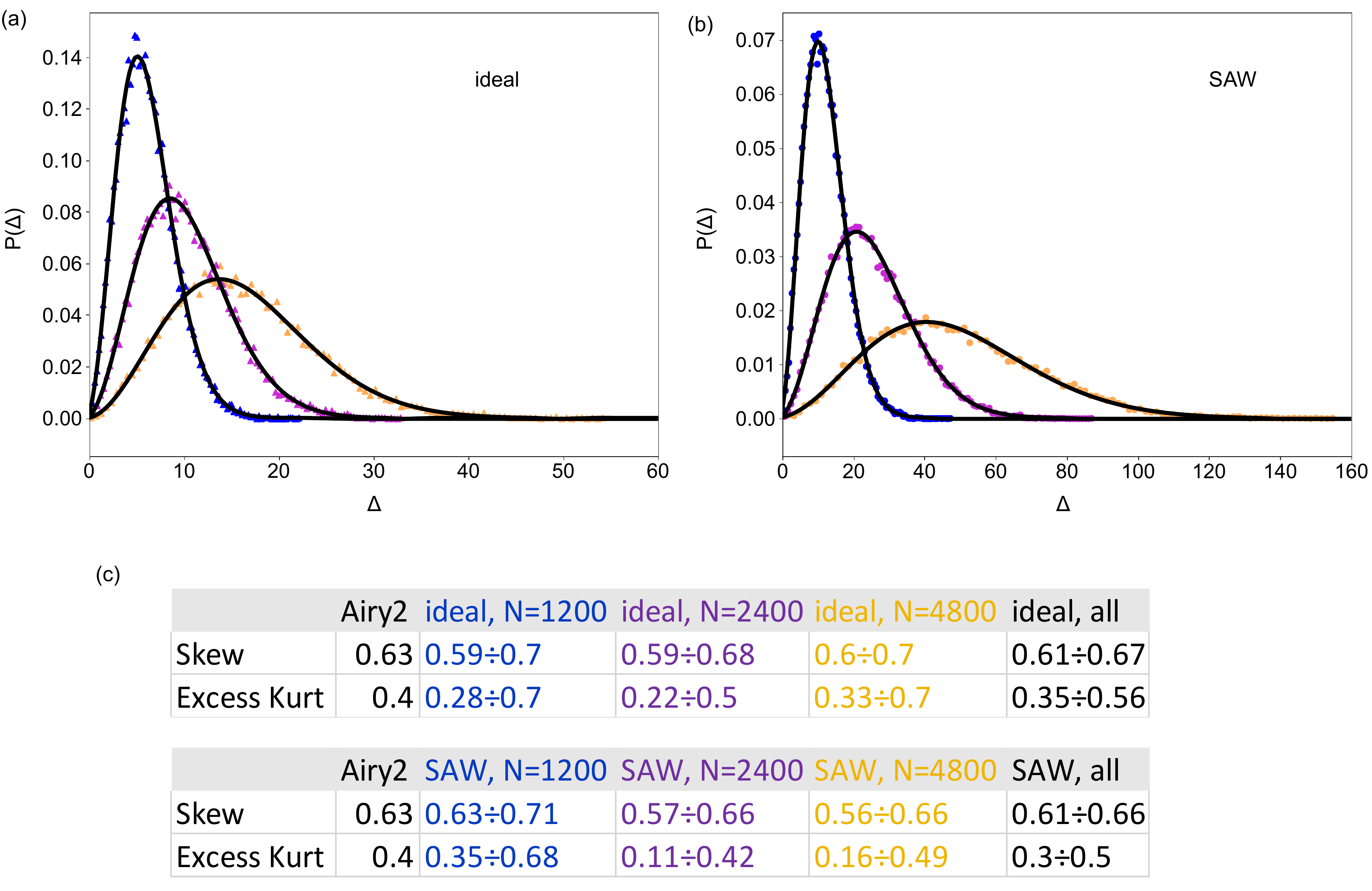}
  \caption{(a),(b): Probability densities of the fluctuations $P(\Delta)$ from simulations versus the theoretical squared Airy function for different chain length $N=1200,2400,4800$ (colors match the colormap used on the graph of the standardized densities in Fig.3(b) of the main text) and (a) ideal and (b) 3D SAW statistics of unrestricted chains. The standardized Airy function with constants \eq{abc} is transformed according to the sample $\mu_\Delta$ and $\sigma_\Delta$ from simulations data according to \eq{rescaled_airy}. (c): The tables with the values of skewness and excess kurtosis for the theoretical PDF, each sample with fixed $N$ and the corresponding moments of the standardized variable $h$ in the merged pool of all values of $N$ analyzed. The confidence intervals for the values of skewness and kurtosis are obtained using bootstrapping and further cut at the 1st and the 99th percentiles.
}
  \label{fig:distrs}
\end{figure*}

Theoretical curves for the original densities, $P(\Delta)$, are straightforward under the inverse transformation
\be
P(\Delta)=c\sigma_\Delta^{-1}\mathrm{Ai}^2\left(a\sigma_\Delta^{-1}(\Delta-\mu_\Delta) + b\right)
\label{rescaled_airy}
\ee
where $\mu_\Delta$ and $\sigma_\Delta$ are the mean and the standard deviation of the fluctuations. In Fig. \ref{fig:distrs} we overlay theoretical densities
$P(\Delta)$ onto the ones from simulations for different values of $N=1200,2400,4800$ and different values of $\nu=1/2$ (ideal) and $\nu \approx 0.588$ (3D self-avoiding walks). Visual inspection allows to infer rather good correspondence between the two. To provide a quantitative comparison we further compute the third and the fourth moments of $h$ (known as skewness, ${\tilde{\mu}}_3=\langle h^3 \rangle$, and kurtosis, ${\tilde{\mu}}_4=\langle h^4 \rangle$; excess kurtosis is ${\tilde{\mu}}_4-3$) for the standardized squared Airy and numerical data (see the inset tables in Fig. \ref{fig:distrs}). The moments for the numerical distributions are computed for each value of $N$, as well as for the merged standardized data with different N (as depicted in the Fig. 3(b) of the main text). The confidence intervals for the values of skewness and kurtosis are obtained using bootstrapping and further cut at the 1st and the 99th percentiles.

\section{Stretching free energy minimization from an ``optimal fluctuation'' perspective}\label{app_bv}

\begin{figure}[ht]
  \centering
  \includegraphics[width=0.4\textwidth]{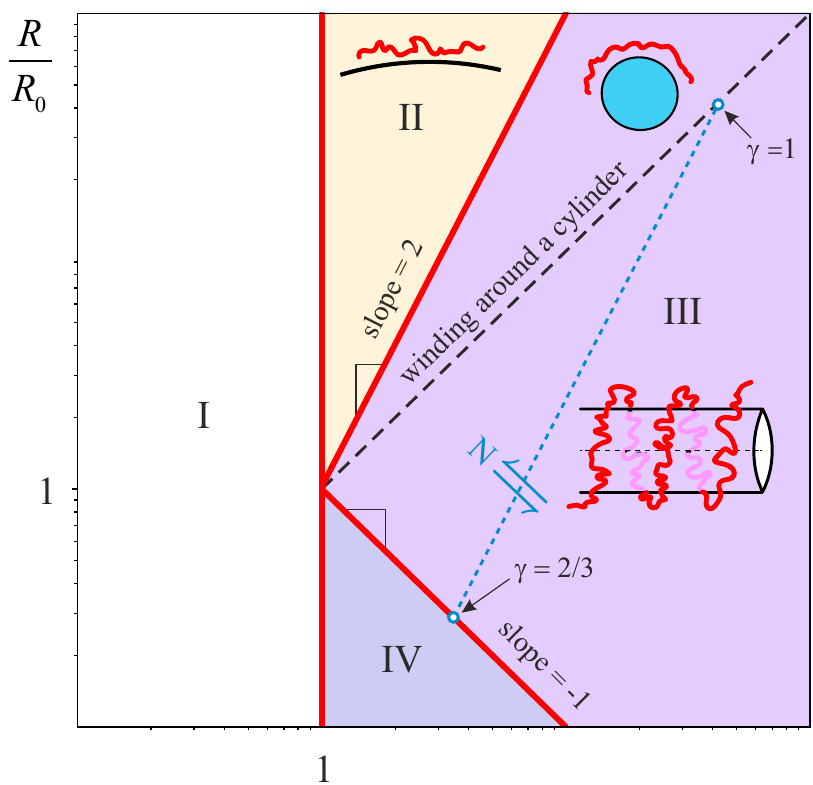}
  \caption{The same diagram as Fig.2(b) in the main text to demonstrate the place of Balagurov-Vaks problem (dashed blue line) in the context of 2D stretched polymer chains. The slopes between the regimes are computed for the particular value of $\nu=1/2$. The arrows correspond to two values of the stretching parameter, $\gamma=2/3$ and $\gamma=1$, between which the mapping to Balagurov-Vaks can be realized for any $N$.}
  \label{fig:diagram_bv}
\end{figure}

The purpose of this section is to establish the connection between our polymer stretching problem and classical Balagurov-Vaks (BV) problem of random walks on the line with randomly distributed traps \cite{balagurov}  (see also later more detailed treatment by Donsker and Varadhan \cite{donsker}).
To see the connection with the stretched polymers, let us return again to formula (4) in the main text and re-write it by assuming, as in the main text, $S = b N^{\gamma}$, with $\gamma < 1$.  Power $\gamma$ can be viewed as a proxy of the distance $S$, characterizing the stretching degree.  Furthermore, we can also say $S = R \theta$, where $\theta$ is the corresponding angle, $\theta<2\pi$ (or $\theta > 2/\pi$) correspond to less than one (or more than one) full turns around the cylinder; in the latter case, $\theta / 2 \pi$ is the winding number. In terms of $\gamma$ and $\theta$, the two terms of free energy read
\be \begin{split}
\frac{F_{\mathrm{circ}}}{k_B T} &  = \frac{\Delta}{R} N^{\frac{\gamma - \nu}{1-\nu}} + \frac{b^2 }{\Delta^2} N^{1+\frac{(1-\gamma )(2\nu-1)}{1-\nu}}  \\ & = N^{\frac{(1-\gamma )(2\nu-1)}{1-\nu}} \left[ \theta N^{-\frac{(1-\gamma)(3 \nu -1 )}{1-\nu}} \frac{\Delta}{b} + \frac{b^2 }{\Delta^2} N \right]  \ .
\label{eq:free_energy_circle_2}
\end{split}
\ee
This result has a transparent connection with BV problem in two cases.  First, if the chain is strongly stretched such that $\gamma \to 1^{-}$, such that $1 - \gamma \ll 1/\ln N$. In that case,
\be
\frac{F_{\mathrm{circ}}}{k_B T}    =  \theta  \frac{\Delta}{b} + \frac{b^2 }{\Delta^2} N  \ .
\label{eq:free_energy_circle_3}
\ee
for arbitrary $\nu$.  Second, if $\nu = 1/2$ and $\theta = N^{1-\gamma}$, in that case
\be
\frac{F_{\mathrm{circ}}}{k_B T}    =   \frac{\Delta}{b} + \frac{b^2 }{\Delta^2} N  \ ;
\label{eq:free_energy_circle_4}
\ee
this latter case corresponds to $\nu = 1/2$ of the polymer. In this case mapping to Balagurov-Vaks is realized along the line $R = b N^{2 \gamma -1}$, which can be also presented as
\be
R/R_0=\frac{1}{\sqrt{N}} \left(S/R_0\right)^2.
\label{eq:bvline}
\ee
The behavior \eq{eq:bvline} is illustrated by the dashed blue line in the $R-S$ diagram, \fig{fig:diagram_bv}.
Interestingly, the slope of this line coincides with the slope of the boundary between flat (II) and curved (III) regimes in the particular case of $\nu=1/2$. However, note that the coefficient in \eq{eq:bvline} is $N$-dependent. Therefore, the mapping to BV in the second case \eq{eq:free_energy_circle_4} might be realized only for particular value of $N$, provided a pair of values $(R/R_0, S/R_0)$ on this diagram.

While in the first case \eq{eq:free_energy_circle_3} the BV mapping is realized in the whole area of the curved regime III, in the second case \eq{eq:free_energy_circle_4} it is not. Along the BV lines of constant $N$ in \fig{fig:diagram_bv} the stretching parameter $\gamma$ changes, and the extremities of these lines provide respective bounds to $\gamma$. As \fig{fig:diagram_bv} suggests, the stretching must be strong enough, $\gamma > 2/3$, otherwise one enters the regime IV of weak fluctuations. On the other hand, it is evident that as long as a BV polymer, being wrapped over cylinder many times, is forced to make a single turn only, $R/R_0 \to S/R_0$, the stretching attains its asymptotic limit, $\gamma \to 1$. This rhymes well with the behavior of the winding number in the second case $\theta = N^{1-\gamma} \gg 1$ for $\gamma<1$. Thus, the region of less than one turn, i.e. between the dashed black line and the boundary II-III, is forbidden for the BV polymers, unless they are fully stretched (the first case).

Let us remind the Balagurov-Vaks setting. Consider an auxiliary 1D problem of random walks on the line with Poisson-distributed absorbing traps. Let $n_{\mathrm{tr}}$ be the average density of traps on the line. Following Balagurov and Vaks \cite{balagurov}, we are interested in the probability $W(N)$ for the walker to survive during ``time'' $N$ (assuming ``diffusivity'' is equal to $b^2/\pi^2$), i.e. with the probability that until time $N$ the walker does not encounter any trap. The probability to have an interval $\Delta$ between nearest neighboring Poisson-distributed traps is equal $\exp(-n_{\mathrm{tr}}\Delta)$. On the other hand, the probability to survive for a ``long time'' $N \gg \Delta^2/b^2$  between absorbing (Dirichlet) boundary conditions on both ends of the interval $\Delta$ is estimated as $\exp(-b^2 N/\Delta^2)$.  The total survival probability is controlled by the Lifshitz's ``optimal fluctuation'' \cite{Lifshitz_Tails}, i.e., by finding such an interval $\Delta$ that maximizes the product of the two above mentioned factors:
\be
W(N) \sim \max_{\Delta} \left[e^{-n_{\mathrm{tr}} \Delta - b^2 N/\Delta^2}\right] \ .
\label{eq:min}
\ee
The connection with \eq{eq:free_energy_circle_3} is now obvious, and $\theta /b$ plays the role of traps density,  $n_{\mathrm{tr}}=\theta/b$.  Clearly,  \eq{eq:free_energy_circle_4} (which is restricted to $\nu = 1/2$ and special value of $R$) corresponds to trap density just $n_{\mathrm{tr}}=1/b$. Note that the derivation of the BV survival probability has relied on the assumption $Nb^2 \gg \Delta^2$, i.e. a the walk between the neighboring traps is constrained. For the case of $\nu=1/2$ this is equivalent to $R/R_0 \ll (S/R_0)^2$ in the polymer problem, which forbids flat geometry. As can be seen from \fig{fig:diagram_bv}, this condition is well satisfied.

Maximization of the expression \eq{eq:min} yields $W(N) \sim \exp \left(-\mathrm{const} \, b^{2/3} n_{\mathrm{tr}}^{2/3} N^{1/3} \right)$, which is exactly the Balagurov-Vaks answer \cite{balagurov} for the 1D survival probability of the unbiased random walk of time $N$ in the Poissonian array of traps.  Due to the analogy, we can call the minus logarithm of the survival probability the ``trap free energy'' \footnote{To be specific, we stick to the first case of strong stretching, $\gamma \to 1$, \eq{eq:free_energy_circle_3}} (dropping from now on the $k_BT$ factor): $-\ln W(N) = F_{\mathrm{trap}} = \theta^{2/3} N^{1/3}$.  The minimal value of the polymer free energy is given by the same formula  $F_{\mathrm{circ}} = \theta^{2/3}N^{1/3} $.

Interestingly, the equivalent to \eq{eq:min} weight was maximized in \cite{Muthukumar:2018} for computation of the correlation function of a polymer chain confined in a gel matrix. In that case the linear term was played by the confinement free energy inside a mesh (generating the exponential distribution of the chain segments lengths), while the quadratic term corresponded to the Rouse relaxation time of each chain segment within the mesh.


In both polymer and BV problem there is in general also the leading extensive term, proportional to $N$.  In BV problem, it is due to a constant bias, $c$, superimposed on the symmetric random walk.  In polymer problem it is a constant energy per every monomer (e.g., a bond energy).  In both cases, therefore,
\be \begin{split}
F_{\mathrm{trap}} & = c N + \left( b n_{\mathrm{tr}} \right)^{2/3} N^{1/3} \ \ \mathrm{and} \\ F_{\mathrm{circ}} & = c N + \theta^{2/3} N^{1/3}  \ , \end{split} \label{eq:identical_free_energies}
\ee
free energies are given by identical expressions, albeit with different physical interpretation of the parameters.

The Legendre transform from $N$ to a conjugate variable, $\lambda$, realized via the inverse Laplace transform of the survival probability $W(N) = \exp \left( - F_{\mathrm{trap}} \right)$ or of the partition sum for a polymer $\exp \left( - F_{\mathrm{circ}} \right)$, gives the spectral density, $\rho(\lambda)$ (see \cite{nieuwenhuizen} for more detail):
\be \begin{split}
\rho(\lambda) & = \frac{1}{2\pi i}\int\limits_{\eps-i\infty}^{\eps+i\infty} e^{-c N - \theta^{2/3} N^{1/3}} \, e^{N\lambda} dN  \\ & \propto \exp \left[ -\theta /\sqrt{
\lambda - c} \right], \quad \lambda>c.
\label{eq:35}
\end{split} \ee

\section{Polymer stretching above a disc in a broader context}\label{app_flow}

\begin{figure*}
\includegraphics[width=0.9\textwidth]{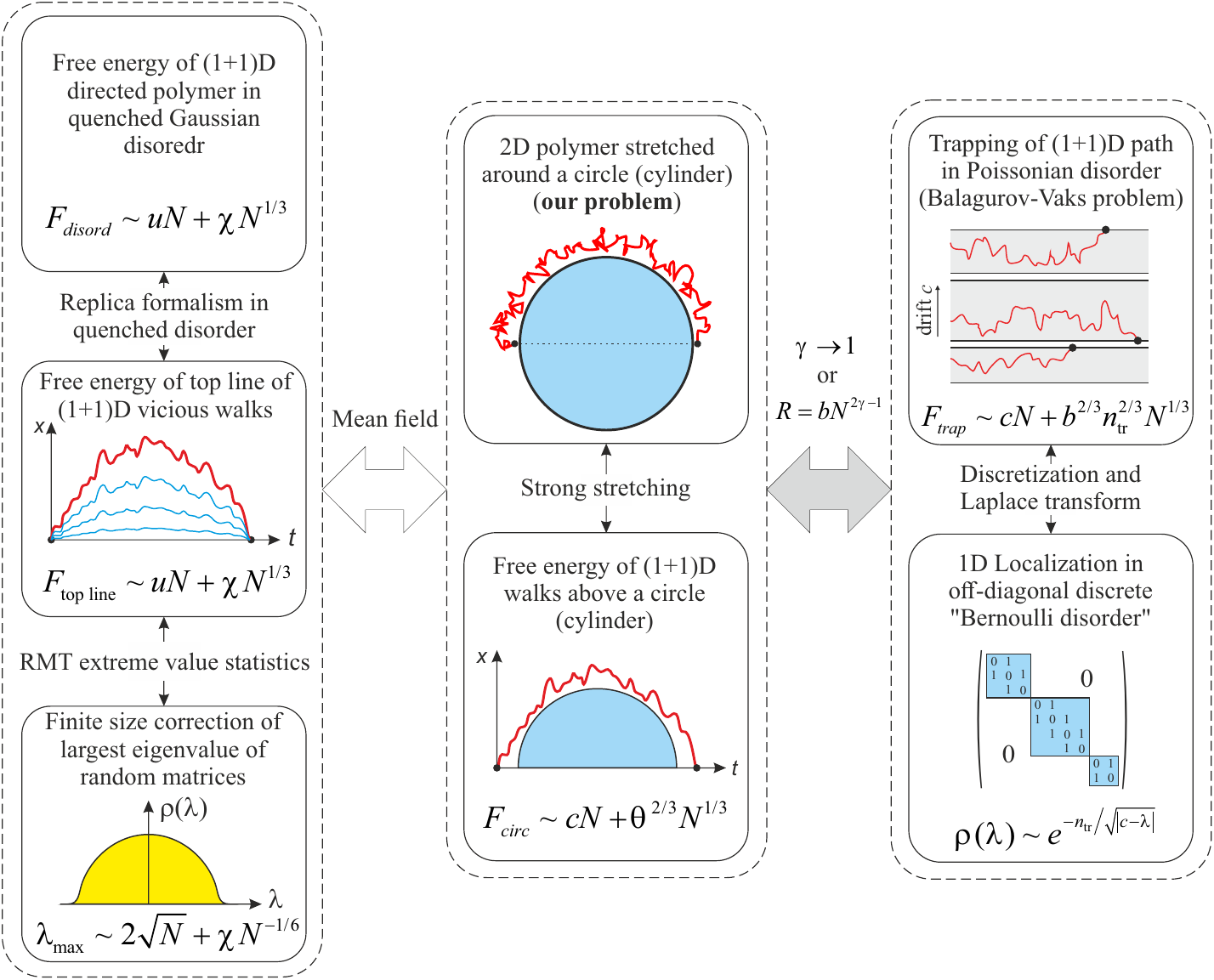}
  \caption{Flowchart of logical connections: place of our ``2D polymer stretching above a curved boundary'' problem in the context of other models and systems in statistical physics.  \textbf{Central column} -- 2D polymer (\textit{top}) and (1+1)D polymer (\textit{bottom}) are equivalent in the strong stretching regime, with free energy $F_{\mathrm{circ}} = c N + \theta^{2/3} N^{1/3}$. \textbf{Right column} -- polymer problem in the proper limit maps onto biased Brownian motion in an array of Poisson distributed traps (\textit{top}), or, equivalently, related to the spectrum of the off-diagonal random Bernoulli matrix (\textit{bottom}).  \textbf{Left column} -- curved polymer stretching problem is a mean field approximation for the top line in the system of (1+1)D vicious (mutually non-intersecting) walks (\textit{center}), which is in turn related to either directed polymer in Gaussian disorder (\textit{top}) and to the maximal eigenvalue statistics in the spectrum of random matrices (\textit{bottom}).  The common motif is the $N^{1/3}$ scaling of the subleading correction term that controls relevant physics in all cases. }
  \label{fig:flowchart}
\end{figure*}

Having established the connection of our polymer problem with that of random walks between traps, we now switch to even further connections to a range of other problems and models.  To facilitate the discussion, we outline it in the flowchart of mutually related ideas presented in Fig. \ref{fig:flowchart}.


The central rectangle in Fig. \ref{fig:flowchart} shows our problem and its limiting regime $\gamma\to 1^{-}$ of strong stretching (as a reminder, stretching of a polymer is characterized by the curvilinear end-to-end distance which we write in terms of $\gamma$ as $S = b N^{\gamma}$).  Minimal value of polymer free energy, as discussed before, is given by $F_{\mathrm{circ}} = c N + \theta^{2/3} N^{1/3}$, where $\theta$ is the winding number, related to the radius of the void, $R = S/ \theta$. The sublinear in $N$ term of free energy represents the curvature-induced finite-size correction.

The right rectangle in the same Fig. \ref{fig:flowchart} depicts the group of problems related to BV model of 1D random walk in the array of Poisson distributed traps, as reviewed in the previous section.  In particular, the bottom panel of the right rectangle schematically depicts the (biased) BV model \cite{balagurov}.  There, we show pictorially a set of randomly positioned traps - thick lines parallel to the time axis.  Within each interval between traps, the walker moves randomly under some constant bias $c$ until it hits one of the boundaries for the first time.  The connectioon to our polymer problem is highlighted by the ``free energy'' expression (\ref{eq:identical_free_energies}), in which trap density is related to the winding angle $\theta$ for the polymer.

In the upper panel on the right hand side we have drawn the typical three-diagonal random matrix with Bernoulli disorder.  Its connection with BV model and, therefore, its relation to our polymer-around-a-cylinder problem can be understood by the following simple calculation.  Let $\rho(\lambda)$ be the spectral density of ensemble of large tridiagonal symmetric matrices, $A_N$, with the bimodal (Bernoulli) distribution of sub-diagonal matrix elements $a_{j,j\pm 1}=\{0,1\}$ as shown below:
\be
A_N = \left(\begin{array}{ccccc}
0 & \eps_1 & 0 & \cdots & 0 \smallskip \\  \eps_1 & 0 & \eps_2 & & \smallskip \\  0 & \eps_2 & 0 & & \smallskip \\ \vdots &  &  &  & \smallskip \\ & & & & \eps_{N-1} \smallskip \\ 0 & & & \eps_{N-1} & 0 \end{array} \right)
\label{e:04c}
\ee
where
\be
\eps_x=\left\{\begin{array}{ll} 1 & \mbox{with probability $p$} \medskip \\
0 & \mbox{with probability $1-p$} \end{array} \right.
\label{e:04d}
\ee
The matrix $A_N$ at each $\eps_x=0$ splits into regular (gapless) three-diagonal ``cage'' of some random size $D$, each can be viewed as a transition matrix of a discrete random walk in the cage $D$. The probability to find such a cage is $Q(D) = p^D$. The spectral density $\rho(\lambda)$ of ensemble of matrices $A_N$ has been exhaustively analyzed in \cite{krapiv,polov} and the tail of $\rho(\lambda)$ near the spectral edge $\lambda\to\lambda_{\max}=2$ reads:
\be
\rho(\lambda)\propto \exp \left[- \frac{\pi \ln p}{\sqrt{|2 - \lambda|}} \right]
\label{e:05}
\ee
Obviously, $\rho(\lambda)$ in \eq{e:05} is the same spectral density as in \eq{eq:35} for properly adjusted drift $c$ and trap density $n_{\mathrm{tr}}$.

Thus, the close similarity between central and right rectangles in the flowchart in Fig. \ref{fig:flowchart} justifies our claim that nontrivial stretched exponent $1/3$ appearing for the random walk or a stretched polymer near the curved boundary points to the intimate connection with stretched exponent for survival probability of (1+1)D trapping problem in the Poissonian disorder.

The left rectangle highlights the known relation between the ground state free energy, $F_{\mathrm{disord}}$ of (1+1)D directed polymer in quenched Gaussian disorder \cite{dotsenko} (upper panel) and the statistics of the top line in the ensemble of (1+1)D ``vicious'' random walks \cite{schehr08} (central panel). Let us note, that the last problem has also the interpretation (after proper rescaling by $\sqrt{N}$) in terms of the largest eigenvalue $\lambda_{\max}$ of the Gaussian ensemble of random matrices. Since the same scaling (subject to numerical factors) is valid for both Gaussian Orthogonal (GOE), and Gaussian Unitary (GUE) ensembles, we do not specify here which particular ensemble is considered. At the spectral edge $\lambda_{\max}$  has the finite-size corrections in $N$ ($N\gg 1$): $\lambda_{\max} = 2\sqrt{N} +\chi N^{-1/6}$, where $\chi$ is $N$-independent and is distributed according to the Tracy-Widom law which takes slightly different forms for GOE and GUE.

The arrow ``Mean field'' designates the mean-field approximation of the many-body system of vicious walks, in which the influence of all trajectories lying below the topmost one, are replaced by the impermeable circular boundary, \cite{spohn_ferrari}. Note that finite-size corrections to the free energies, $F_{\mathrm{disord}}$ and $F_{\mathrm{upper}}$, have the same scaling as the one for $F_{\mathrm{circ}}$: in all cases the corresponding finite-size sublinear in $N$ terms are of order of $N^{1/3}$.

We should emphasize that the above mentioned similarity, however attractive, is not complete.  Although valid on averages, it cannot be extended on distributions: the partition functions of a polymer in quenched Gaussian disorder and fluctuations of the topmost vicious walks have the Tracy-Widom distribution, while the constrained random walk above the boundary is given by squared Airy function \cite{spohn_ferrari,valov_fixman}. Here we report the equivalent squared Airy PDF of fluctuations in the stretched polymer problem at various degrees of stretching. Apparently, this difference in distribution is the consequence of the fact that we have replaced the true many-body system (such as vicious walks) by its one-body mean-field analog.

To summarize, scaling analysis of a polymer strongly stretched around a cylinder reveals an unusual behavior of free energy $F_{\mathrm{circ}} \sim cN + \theta^{2/3}N^{1/3}$ that points to an array of deep connections with a variety of problems in equilibrium and non-equilibrium statistical physics and random matrix theory, ranging from KPZ to Balagurov-Vaks problem, Lifshitz tails, Andresen localization, vicious random walks, etc.  Although some of the arguments in the last section have intentionally tentative, hypothetical, and sometimes even speculative character, it seems to us that together they paint an exciting picture.

\bibliography{ABC}





\end{document}